# Multi-Terminal Memtransistors from Polycrystalline Monolayer MoS$_2$

*Vinod K. Sangwan,[1,*] Hong-Sub Lee,[1,*] Hadallia Bergeron,[1] Itamar Balla,[1] Megan E. Beck,[1] Kan-Sheng Chen,[1] and Mark C. Hersam[1,2,3,#]*

[1]Department of Materials Science and Engineering, Northwestern University, Evanston, Illinois 60208, USA.

[2]Department of Chemistry, Northwestern University, Evanston, Illinois 60208, USA.

[3]Department of Electrical Engineering and Computer Science, Northwestern University, Evanston, Illinois 60208, USA.

*These authors contributed equally.

#E-mail: m-hersam@northwestern.edu

In the last decade, a 2-terminal passive circuit element called a memristor has been developed for non-volatile resistive random access memory and has more recently shown promise for neuromorphic computing.[1, 2, 3, 4, 5, 6] Compared to flash memory, memristors have higher endurance, multi-bit data storage, and faster read/write times.[4, 7, 8] However, although 2-terminal memristors have demonstrated basic neural functions, synapses in the human brain outnumber neurons by more than a factor of 1000, which implies that multi-terminal memristors are needed to perform complex functions such as heterosynaptic plasticity.[3, 9, 10, 11, 12, 13] Previous attempts to move beyond 2-terminal memristors include the 3-terminal Widrow-Hoff memistor[14] and field-effect transistors with nanoionic gates[15] or floating gates,[16] albeit without memristive switching in the transistor.[17] Here, we report the scalable experimental realization of a multi-terminal hybrid memristor and transistor (i.e., memtransistor) using polycrystalline monolayer MoS$_2$. Two-dimensional (2D) MoS$_2$ memtransistors show gate tunability in individual states by 4 orders of magnitude in addition **to large switching ratios with high cycling endurance and long-term retention of states. In addition to conventional neural learning behavior of long-term potentiation/depression, 6-terminal MoS$_2$ memtransistors possess gate-tunable heterosynaptic functionality that is not achievable using 2-terminal memristors. For example, the conductance between a pair of two floating electrodes (pre-synaptic and post-synaptic neurons) is varied by ~10× by applying voltage pulses to modulatory terminals. *In situ* scanning probe microscopy, cryogenic charge transport measurements, and device modeling reveal that bias-induced MoS$_2$ defect motion drives resistive switching by dynamically varying Schottky barrier heights. Overall, the seamless integration of a memristor and transistor into one multi-terminal device has the potential to enable complex Hebbian learning in addition to providing opportunities for studying the unique physics of defect kinetics in 2D materials.[18, 19, 20, 21, 22]**

Uniform polycrystalline monolayer MoS$_2$ films with an average grain size of 3–5 μm were grown by chemical vapor deposition (CVD) on SiO$_2$/Si substrates (Methods) and characterized using X-ray photoelectron, photoluminescence, and Raman spectroscopies (Figs. 1a–c and Extended Data Fig. 1). MoS$_2$ memtransistors were fabricated in a field-effect geometry with channel lengths ($L$) and widths ($W$) varying from 5 μm to 150 μm (Figs. 1d–f). Since clean interfaces between MoS$_2$ channels and metal electrodes were found to be critical, an unconventional photolithography process was developed based on a polymethylglutarimide (PMGI) and photoresist bilayer (Methods and Extended Data Fig. 2).

At large drain bias $V_D$ in the sub-threshold regime (gate bias $V_G$ < threshold voltage $V_{th}$), a typical memtransistor is in a high resistance state (HRS) for the $V_D$ sweep from 0 to 80 V (sweep 1) and gradually changes to a low resistance state (LRS) (Fig. 2a). The device maintains LRS during the sweep from 80 to 0 V (sweep 2), resets to HRS during the sweep from 0 to –80 V (sweep 3), and maintains HRS from –80 to 0 V (sweep 4). Thus, these devices act as LRS-HRS memtransistors. Fig. 2b shows that when $V_G$ is varied from 50 to –50 V, both LRS and HRS resistances change by a factor of ~$10^4$ and the switching ratio ($I_{LRS}/I_{HRS}$ at $V_D = 0.5$ V) is tuned from 300 to 8 (inset Fig. 2e). Due to the n-type MoS$_2$ channel, the forward-biased ($V_D > 0$ V) device is completely OFF at $V_G = -50$ V (Fig. 2c), and the reverse-biased ($V_D < 0$ V) device is insulating for a range of $V_D$ depending on the applied $V_G$ (Fig. 2b and Extended Data Fig. 3c). The gate leakage ($I_G$–$V_G$) remains below 200 pA during high $V_D$ and $V_G$ sweeps (Extended Data Fig. 3a).

Unlike filament-based resistive switching, MoS$_2$ memtransistors do not require an electroforming process, although the switching ratio increases with increasing $V_D$ sweep range (Extended Data Fig. 3b). The largest switching ratio of >$10^2$ was obtained from devices with $W = 100$–$150$ μm and $L = 5$–$15$ μm. The devices show bipolar resistive switching where reversing the bias polarity is essential to restoring the initial resistance states (Extended Data Figs. 3d,e). The hysteresis in the $I_D$–$V_D$ characteristics of these MoS$_2$ memtransistors is fundamentally distinct from commonly reported hysteresis in $I_D$–$V_G$ curves due to oxide-related traps in field-effect transistors.[23, 24] Instead, the transfer characteristics ($I_D$–$V_G$) of MoS$_2$ memtransistors in LRS and HRS show significant shifts (~10 V) in threshold voltage ($V_{th}$) with curves that cross each other at $V_G = V_{cross}$ (Fig. 2c). HRS shows up to 100× higher resistance for $V_G < V_{cross}$ and 2× higher field-effect mobility ($\mu$) at $V_G > V_{cross}$ compared to LRS. Therefore, the forward bias switching loop changes from counterclockwise (LRS to HRS) for $V_G < V_{cross}$ to clockwise (HRS to LRS) for $V_G > V_{cross}$ (Fig. 2b and Extended Data Fig. 3f).

Fig. 2d shows endurance characteristics for a MoS$_2$ memtransistor that was switched 475 times between LRS and HRS through full sweep cycles. Within a subset, $I_D$ saturates to an upper value via stretched exponentials (Extended Data Figs. 4 and 5a). Between neighboring subsets, $I_D$ jumps randomly to a ~10× smaller value followed by the same inverse exponential growth, which suggests an oxide-related trap-release process activated by large fields near the source electrode in forward bias.[23] This behavior is reduced under reverse-bias (Extended Data Fig. 5b), possibly due to smaller band-bending near the drain electrode in reverse bias. Since both HRS and LRS show similar transitory decays, the switching ratio (~100) remains relatively constant (Fig. 2d). Individual resistance states measured up to 24 hours show a projected retention of distinct states for timescales on the order of years (Fig. 2e and Extended Data Fig. 5e). A statistical study of 62 devices that were fabricated with identical geometry on a single chip showed a log-normal distribution of switching ratios (Fig. 2f). Device-to-device variability is attributed to spatial inhomogeneity in the CVD $MoS_2$ film.

Due to its 2D nature, the $MoS_2$ channel allows the switching mechanism to be probed via *in situ* electrostatic force microscopy (EFM) as shown in Figs. 3a,b. The drain contact and conductive cantilever tip were biased independently (Extended Data Fig. 6), while the cantilever phase shift was recorded as the tip traced the topography profile 50 nm above the surface. Since the phase shift is proportional to the square of the potential difference between the tip and the underlying surface, the phase image provides a map of the local potential.[20] Line profiles reveal a sharper potential drop (larger field) at the source in forward HRS compared to forward LRS, suggesting a larger contact resistance in HRS than LRS (Fig. 3b). Reverse HRS also shows a larger field at the drain compared to reverse LRS, although this difference is smaller, consistent with the smaller switching ratio at reverse bias. Multiple line profiles confirm that the differences in HRS and LRS electric fields are consistent along the entire channel width, despite grain boundary induced EFM phase variations away from the contacts (Extended Data Fig. 7). Since a reverse-biased Schottky diode at the source (drain) dominates $I_D$ at $V_D > 0$ ($V_D < 0$), EFM measurements provide direct evidence that the HRS to LRS (LRS to HRS) switching in memtransistors is caused by dynamic tuning of Schottky barriers. Using variable temperature ($T$) charge transport measurements and assuming a 2D thermionic emission model, the effective barrier heights ($\Phi_b$) in different states are extracted from the slope of $\ln(I/T^{3/2})$ versus $1/T$ at different $V_G$ (Extended Data Fig. 8). Thermionic emission dominates at $V_G < V_{FB}$ (flat-band voltage) producing linearly varying $\Phi_b$ with $V_G$. Thermally-assisted tunneling through a deformed Schottky barrier begins to contribute at $V_G > V_{FB}$, resulting in deviation from the linear trend (Fig. 3c). Thus, the Schottky barrier height ($\Phi_{sb} = \Phi_b$) is extracted from $V_G = V_{FB}$. $V_{FB}$ decreases from 2 V (HRS) to –8 V (LRS), consistent with a $V_{th}$ shift of 15 V for the same device (Extended Data Fig. 8). In addition, $\Phi_{sb}$ decreases from 125 meV (HRS) to 80 meV (LRS), confirming the observations from EFM.

Charge transport in MoS$_2$ LRS-HRS memtransistors can be described by Schottky barrier transistors (equations 1 and 2 below) where $\Phi_b$ is a function an internal state variable ($w$) defined as the width of the region with excess dopants ($\Delta n$) (Extended Data Fig. 9a).[1, 17, 25] Tuning the Schottky barrier through increased doping near contacts ($\Phi_b \sim \sqrt{\Delta n . w}$) is a standard practice in conventional field-effect transistors.[26, 27] Similarly, we propose that defects in MoS$_2$ act as dopants with their local migration under an applied bias being facilitated by grain boundaries, as previously observed under transmission electron microscopy and explained by *ab initio* calculations.[18, 19, 21] Thus, we develop a memtransistor model where the effective $\Phi_b$ changes by image charge lowering and tunneling at high biases (see Methods), resulting in the following coupled equations:

$$I_D = D \exp\left(\frac{e(V_G - V_{th})}{c_r . k_B T}\right)\left(1 - \exp\left(-\frac{eV_D}{k_B T}\right)\right) \exp\left(\frac{\Phi_b(w)}{k_B T}\right) \quad (1)$$

$$\Phi_b(w) = \phi_{b0} - \frac{e}{\varepsilon_s}\sqrt{\frac{\Delta n.w}{4\pi}} + \sqrt{\frac{e}{4\pi\varepsilon_s}}\left(\frac{2en(\phi_{b0}+A|V_d|)}{\varepsilon_s}\right)^{1/4} \quad (2)$$

$$\frac{\partial w}{\partial t} = E.I_D\{1 - [(w-0.5)^2 + 0.75]^p\} \quad (3)$$

where A, D, E, and $\phi_{b0}$ are fitting parameters, and *e*, *n*, $\varepsilon_s$ are electronic charge, doping level, and dielectric constant of monolayer MoS$_2$, respectively. Equation (3) captures the non-linear kinetics of the dopants near the contacts using a window function where the degree of non-linearity is defined by a positive integer *p* (see Methods).[28] This model agrees well with the experimental data (Fig. 2a and Extended Data Fig. 9) and helps establish that the dependence of *w* on $V_G$ (through $I_D$) embodies the most essential feature of the MoS$_2$ memtransistor and distinguishes it from a 2-terminal memristor.[17] To further illustrate the non-trivial switching, we fabricated MoS$_2$ devices without the PMGI-based process and used residual ~1.5 nm thick photoresist layer as a tunneling barrier between MoS$_2$ and the metal contacts (Figs. 3d,e and Extended Data Figs. 10a-d). The tunneling barrier de-pins the Fermi level and causes the resistance states to change abruptly upon crossing 0 V, resulting in LRS-LRS memtransistors. Overall, the switching mechanism of both LRS-HRS and LRS-LRS memtransistors can be described by two memristors at the contacts connected by a field-effect transistor (see Supplementary notes on switching mechanism and Extended Data Fig. 10e).

While grain boundaries in polycrystalline MoS$_2$ memtransistors enable large switching ratios and prevent electrical breakdown by lowering Schottky barriers through dynamic defect migration, control devices on single grains of CVD-grown MoS$_2$ show a qualitatively different reversible breakdown phenomenon (Fig. 4a). This breakdown event is marked by a sharp drop in the conductance at a voltage $V_{br}$ during sweep 1 that increases again to the original value during sweep 2. This abrupt change in charge transport is accompanied by the emergence of dendritic features (~300–500 nm) close to the source (upper inset in Fig. 4a and Extended Data Figs. 11a,b). Control devices reveal a linear correlation of breakdown current ($I_{br}$) with the width of the source electrode ($W_{src}$) and no correlation with $L$ (lower inset in Fig. 4a and Extended Data Fig. 12). Since $I_D$ normally scales as $W/L$, the observed deviation in this scaling at breakdown suggests that the reverse-biased Schottky diode at the source electrode creates a bottleneck for electron injection at the channel,[27] ultimately causing electromigration at the source contact.[7] This reversible breakdown in CVD $MoS_2$ contrasts the irreversible breakdown in exfoliated $MoS_2$ by Joule heating.[29] In the high bias limit (±120 V), polycrystalline $MoS_2$ memtransistors also degrade irreversibly in a manner that shows light emission during each subsequent sweep (Extended Data Figs. 11c-e).

The 2D planar geometry of the $MoS_2$ memtransistor provides a platform to realize multi-terminal neural circuits mimicking multiple synaptic connections in neurons. For example, in a 6-terminal memtransistor, the conductance between any permutation of two of the four inner electrodes can be modulated by high-bias pulses applied to the outer two electrodes while the inner electrodes are kept floating (Fig. 4b and Extended Data Fig. 13). To achieve heterosynaptic plasticity, the conductance between pre-synaptic and post-synaptic neurons should be controlled by additional modulatory terminals.[30] While this type of modulation has been shown previously in Ag-based cationic memristors, this design is limited to only 3 terminals due to the requirement of filament formation across the channel.[6, 30] In contrast, the $MoS_2$ memtransistor enables larger numbers of terminals and facile tuning via modulation of the Schottky barrier localized at each terminal (see Supplementary notes on switching mechanism). Multi-terminal memtransistors also allow further tuning of heterosynaptic plasticity via a gate electrode where the switching ratio between any permutation of side electrodes can be controlled from ~2 to 10-fold by varying $V_G$ from 50 to 20 V (Extended Data Fig. 13f).

MoS$_2$ memtransistors also demonstrate long-term potentiation and long-term depression that mimic excitatory and inhibitory synapses in organisms (Fig. 4c). Post-synaptic current (PSC) is shown to increase and decreases exponentially with repetition of positive and negative bias pulses of 1 msec duration (Fig. 4c). The linearity of pulse training is comparable to metal-oxide memristors and can be further improved by employing bipolar pulsing schemes.[11, 31] Mimicking spike-timing dependent plasticity (STDP), paired pulses separated by a time interval ($\Delta t$) induce positive and negative changes in synaptic weight with positive and negative pulses (Fig. 4d), resulting in time constants of ~2 and 6 msec, respectively, which are comparable to biological synapse response times.[11, 32]

In conclusion, MoS$_2$ memtransistors combine resistive switching with transistor gating to realize nonlinear charge transport with wide tunability of individual states and switching ratios. In contrast to devices that require single-crystal MoS$_2$ flakes, the utilization of polycrystalline MoS$_2$ films allows straightforward scaling of this technology to large-area device arrays and integrated circuits. The 2D planar geometry of monolayer MoS$_2$ further enables the realization of multi-terminal memtransistors with unprecedented heterosynaptic plasticity that holds promise for complex Hebbian learning from multiple inputs in neuromorphic computing by mimicking biological neurons with multiple synapses.

**Online Content:** Methods, along with any additional Extended Data display items, are available in the online version of the paper; references unique to these sections appear only in the online paper.

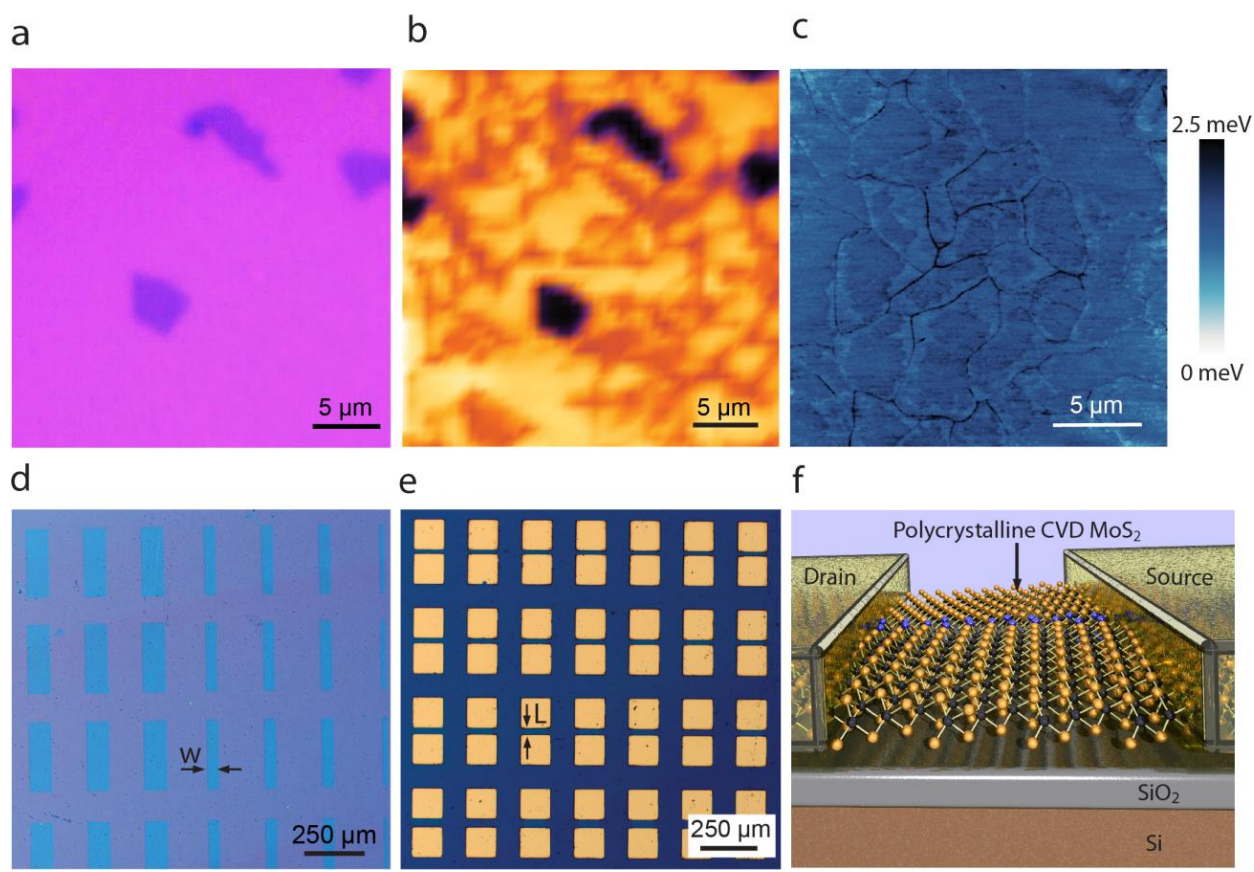

**Figure 1: Architecture of the MoS$_2$ memtransistor. a**, Optical micrograph of CVD-grown polycrystalline monolayer MoS$_2$. The darker regions are from sparse regions of bilayer MoS$_2$. **b**, Spatial mapping of photoluminescence intensity (wavelength = 670 nm) for the area shown in (a). The darkest regions correspond to bilayer MoS$_2$, while the additional features correspond to grain boundaries within monolayer MoS$_2$. **c**, Lateral force microscopy retrace image of monolayer MoS$_2$ showing grain boundaries (see Extended Data Fig. 1e for topography image). **d**, Optical micrograph of an array of MoS$_2$ monolayer strips with varying width (*W*), etched by reactive ion etching before metallization. **e**, Optical micrograph of fully fabricated MoS$_2$ memtransistors with varying channel length (*L*). **f**, Schematic of the MoS$_2$ memtransistor device on 300 nm thick thermal SiO$_2$ on doped Si (gate).

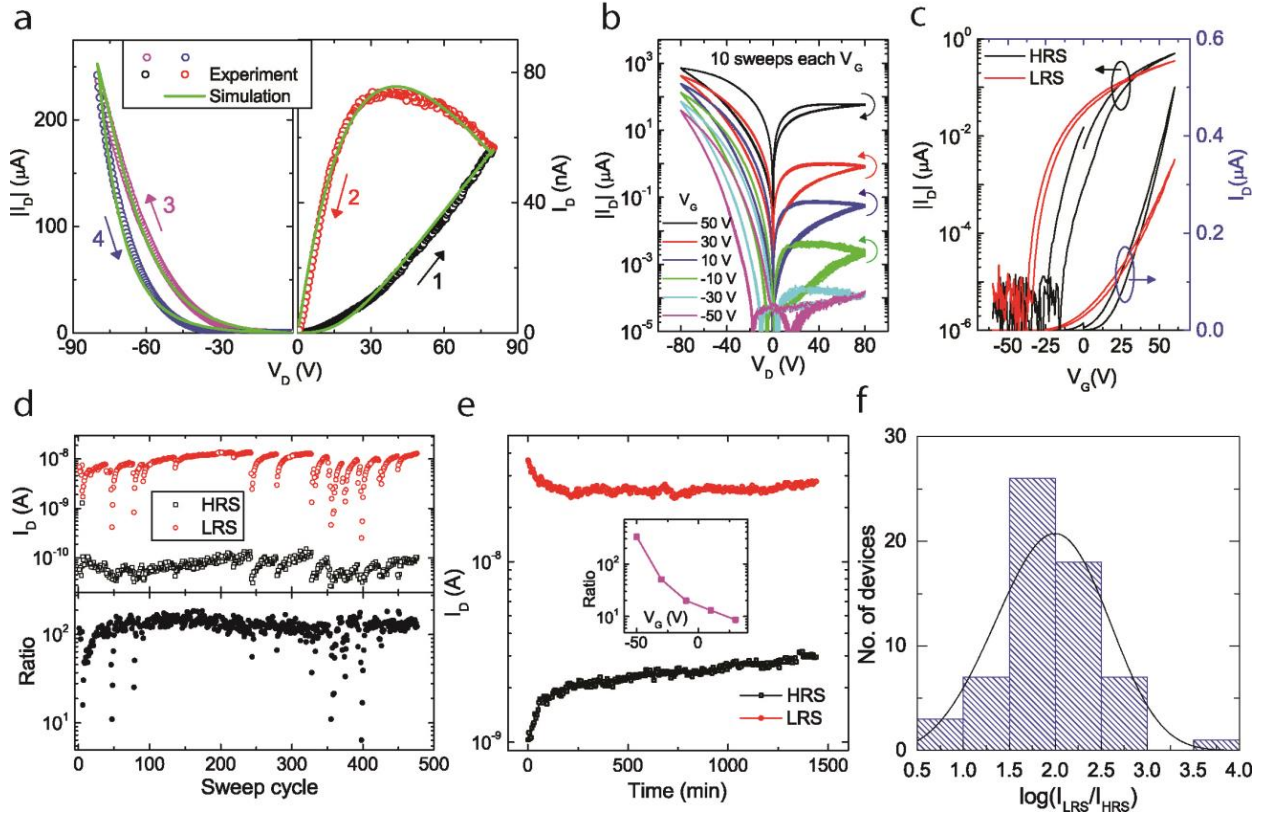

**Figure 2: Charge transport characteristics, endurance, retention, and statistics of MoS$_2$ memtransistors. a**, *I-V* characteristics (open circles) of a MoS$_2$ memtransistor (*L* = 5 µm, *W* = 100 µm) at a gate bias $V_G$ = 10 V. The direction of the drain bias ($V_D$) sweep 1 → 4 is highlighted with colored arrows. The sweep rate is 10 V/s throughout unless mentioned otherwise. The solid green line is simulated data from a memtransistor model (see Methods and Extended Data Fig. 9). **b**, *I-V* characteristics showing 10 consecutive sweeps at each gate bias $V_G$ for the same device where the switching directions are shown by the curved arrows. $V_G$ was stepped down between each $V_D$ sweep cycle. **c**, Transfer characteristics of a memtransistor at $V_D$ = 0.1 V showing a shift

in threshold voltage ($V_{th}$) and field-effect mobility ($\mu$) from $V_{th} = 20$ V and $\mu \sim 0.6$ cm$^2$/Vs in HRS to $V_{th} = 10$ V and $\mu \sim 0.3$ cm$^2$/Vs in LRS. The curves cross at $V_G = V_{cross} \sim 30$ V. **d**, Endurance of current (top) and on/off ratio (bottom) characteristics at $V_D = 0.5$ V in HRS (sweep 1) and LRS (sweep 2) states for 475 consecutive sweep cycles of a memristor at $V_G = 40$ V (see Extended Data Fig. 4 for all $I_D$-$V_D$ curves). **e**, Retention of HRS and LRS currents measured at $V_D = 100$ mV and $V_G = 0$ V for 24 hours. The inset shows the gate-tunability of the switching ratio. **f**, Histogram of the largest on/off ratio of 62 distinct memtransistors ($L = 5$ µm, $W = 100$ µm), where the distribution is fit with a log-normal curve with mean and variance in log($I_{LRS}/I_{HRS}$) of 1.96 and 0.54, respectively.

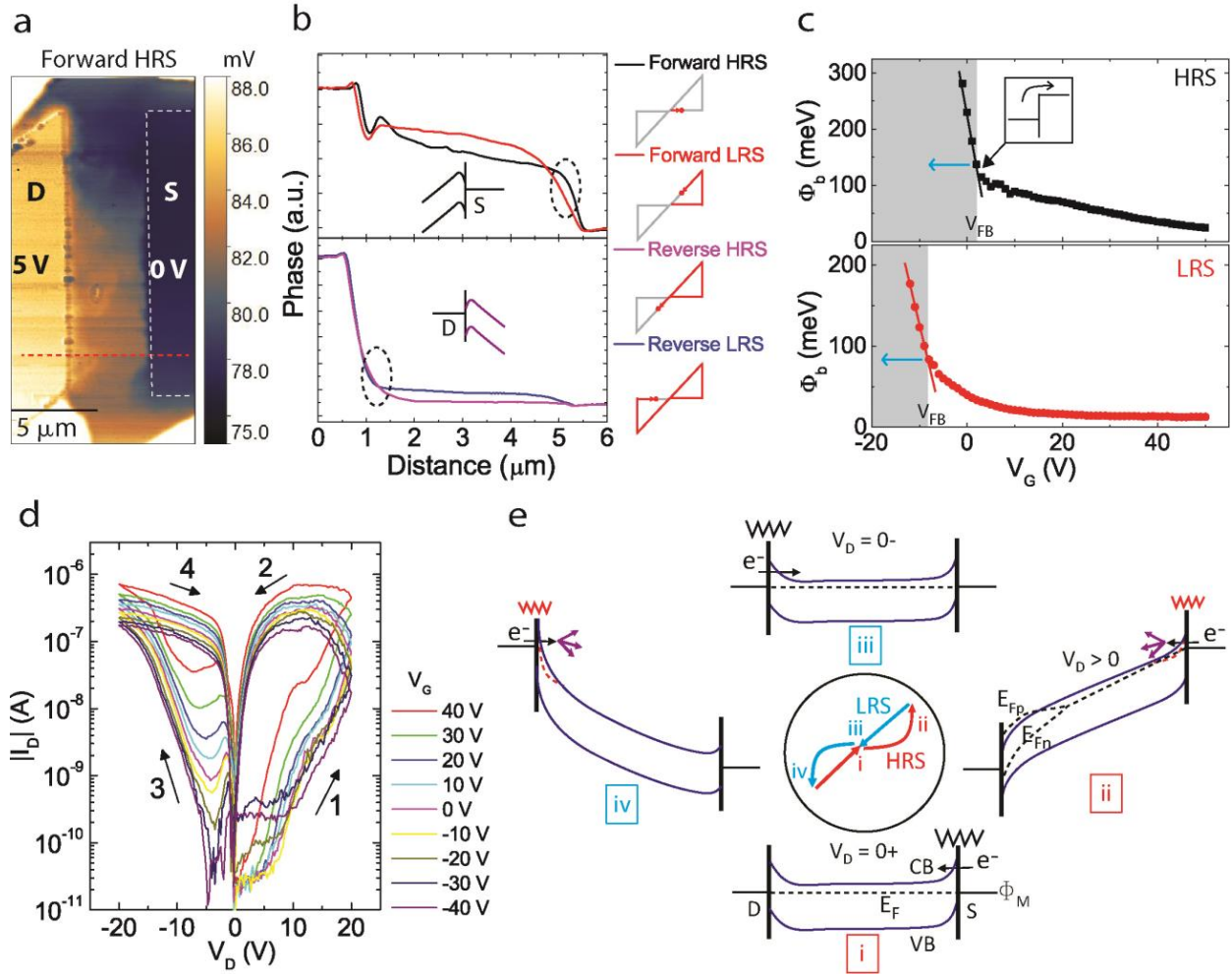

**Figure 3:** *In situ* measurements and switching mechanism. **a**, EFM phase image of a MoS$_2$ memtransistor at $V_D$ = 5 V, $V_S$ = 0 V, and $V_{tip}$ = 0 V in forward HRS. **b**, Line profiles of EFM phase along the red dashed lines in (a) for EFM images taken at forward HRS, forward LRS, reverse HRS, and reverse LRS, respectively, as shown schematically on the right (see Extended Data Figs. 6c-f for EFM images). Dashed ovals show band-bending near the source electrode in forward HRS and LRS and near the drain in reverse LRS and HRS. The dip near the drain in the top panel is a topography artifact. **c**, $V_G$-dependent effective Schottky barrier height ($\Phi_b$) extracted from variable temperature conductivity measurements in HRS (top) and LRS (bottom). $V_{FB}$ is a value of $V_G$ where $\Phi_b$ deviates from the linear dependence of the thermionic emission model. The inset shows the flat band condition at $V_G = V_{FB} = 2$ V, $\Phi_b = 125$ meV in HRS and $V_{FB} = -8$ V, $\Phi_b = 80$ meV in LRS (see Methods and Extended Data Fig. 8 for details). **d**, *I-V* characteristics of a LRS-LRS MoS$_2$ memtransistor with a ~1.5 nm thick photoresist layer under the metal contact at different $V_G$. The direction of the $V_D$ sweep 1 → 4 is highlighted by arrows. $V_G$ was stepped down between each $V_D$ sweep cycle. **e**, Schematic showing the energy band diagram of a LRS-LRS memtransistor at the 4 stages of switching shown in the center. $E_{fp}$ and $E_{fn}$ are non-equilibrium quasi-Fermi levels for holes and electrons, respectively (see Methods and Extended Data Fig. 10 for details).

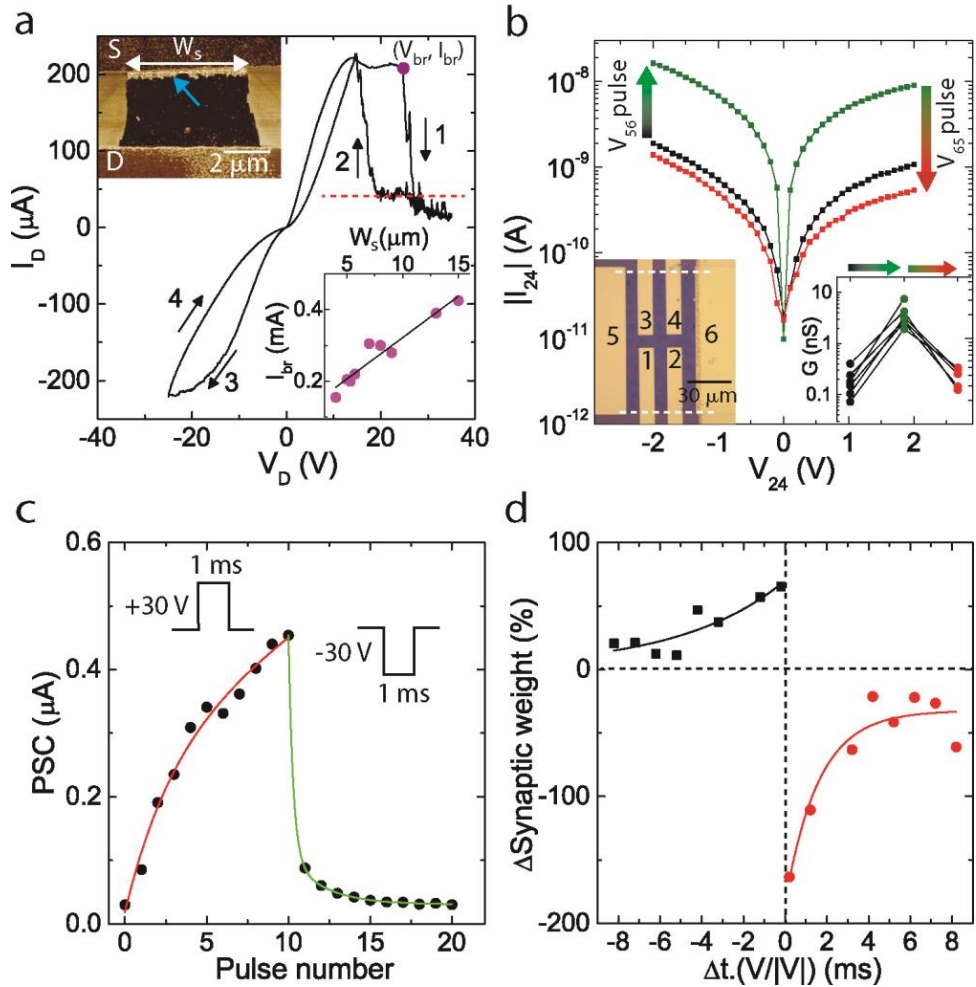

**Figure 4: Control devices and neural functions of memtransistors. a**, $I$-$V$ characteristics (at $V_G$ = 60 V) of a control device on an individual monolayer $MoS_2$ domain without grain boundaries showing breakdown. The breakdown voltage ($V_{br}$) and breakdown current ($I_{br}$) are defined by the point preceding the sharp downward transition in sweep 1. The upper inset shows an AFM phase image of a device with degraded material near the source electrode (arrow). The lower inset shows a linear relation between $I_{br}$ and the width of source contact $W_s$ for 9 devices with different flake shapes (see Extended Data Figs. 11,12). **b**, $I_{24}$-$V_{24}$ characteristics between terminal 2 and 4 of a 6-terminal $MoS_2$ memtransistor (left inset) at three stages of pulses at constant $V_G$ = 20 V: (1) black: before any pulse; (2) green: after applying four pulses of –80 V between terminals 5 and 6 while terminals 1–4 were floating; (3) red: after applying three pulses of –80 V between terminals 6 (drain) and 5 (ground) while terminals 1–4 were floating. The white dashed lines in the left inset show the edges of the patterned $MoS_2$. The right inset shows heterosynaptic plasticity by ~10-fold reversible changes in conductance ($G = I_{ij}/V_{ij}$) between all six combinations of four terminals, ij= 12, 13, 14, 23, 24, and 34 (see Extended Data Fig. 13). **c**, Post-synaptic current (PSC) versus pulse number showing long-term potentiation and depression with 30 V and –30 V pulses, respectively. The solid lines show biexponential fits. **d**, Measured change in synaptic weight change (normalized to maximum weight) with relative time intervals between paired pulses of 40 V and –40 V. The solid lines are exponential fits with time constants of 1.6 and 5.5 msec for positive and negative pulses, respectively. $V_G = 0$ V throughout all measurements in (c) and (d).

ACKNOWLEDGMENTS

This research was supported by the Materials Research Science and Engineering Center (MRSEC) of Northwestern University (NSF DMR-1720139) and the 2-DARE program (NSF EFRI-1433510). CVD growth of $MoS_2$ was supported by the National Institute of Standards and Technology (NIST CHiMaD 70NANB14H012). Charge transport instrumentation was funded by an ONR DURIP grant (ONR N00014-16-1-3179). H.-S.L. also acknowledges the Basic Science Research Program of the National Research Foundation of Korea (NRF) that is funded by the Ministry of Education (2017R1A6A3A03008332). H.B. acknowledges support from the NSERC Postgraduate Scholarship-Doctoral Program. H.B. and M.E.B. acknowledge support from the National Science Foundation Graduate Research Fellowship. This work made use of the Northwestern University NUANCE Center and the Northwestern University Micro/Nano Fabrication Facility (NUFAB), which are partially supported by the Soft and Hybrid Nanotechnology Experimental (SHyNE) Resource (NSF ECCS-1542205), the Materials Research Science and Engineering Center (NSF DMR-1720139), the State of Illinois, and Northwestern University. The authors thank Dr. J. J. McMorrow for assistance with photolithography, Xiaolong Liu for assistance with lateral force microscopy, and Sina Mohseni for assistance with atomic force microscopy.

AUTHOR CONTRIBUTIONS

V.K.S., H.S.L., and M.C.H. conceived the idea and designed the experiments. V.K.S. and H.S.L. fabricated, measured, and analyzed all the devices. H.B. and I.B. conducted growth of $MoS_2$ and materials characterization. V.K.S. developed the memtransistor model. M.B. assisted in model fitting and device fabrication. K.-S.C assisted in electrostatic force microscopy. All authors wrote the manuscript and discussed the results at all stages.

AUTHOR INFORMATION

Reprints and permissions information is available at www.nature.com/reprints. The authors declare no competing financial interests. Readers are welcome to comment on the online version of the paper. Publisher's note: Springer Nature remains neutral with regard to jurisdictional claims in published maps and institutional affiliations. Correspondence and requests for materials should be addressed to M.C.H. (m-hersam@northwestern.edu).